\documentclass[aps,prl,twocolumn,superscriptaddress,floatfix]{revtex4}
\usepackage{graphicx}
\usepackage{amssymb}

\def\LFAOFx{LaFeAsO$_{1-x}$F$_x$}
\def\PFAOFx{PrFeAsO$_{1-x}$F$_x$}
\def\NFAOFx{NdFeAsO$_{1-x}$F$_x$}
\def\SFAOFx{SmFeAsO$_{1-x}$F$_x$}
\def\RFAOFx{\textit{R}FeAsO$_{1-x}$F$_x$}
\def\RFAOF{\textit{R}FeAsO$_{1-x}$F$_x$}
\def\PFAOF{PrFeAsO$_{0.87}$F$_{0.13}$}
\def\NFAOF{NdFeAsO$_{0.85}$F$_{0.15}$}
\def\LFAOF{LaFeAsO$_{0.85}$F$_{0.15}$}
\def\Tc{T$_\mathrm{c}$}
\def\TN{T$_\mathrm{N}$}

\begin{document}

\title{Spatial Inhomogeneity in \RFAOF ($R=$Pr, Nd) Determined from Rare Earth Crystal Field Excitations}

\author{E. A. Goremychkin}
\affiliation{Argonne National Laboratory, Argonne, IL 60439, USA}
\affiliation{ISIS Facility, Rutherford Appleton Laboratory, Chilton, Didcot, Oxon OX11 OQX, United Kingdom}
\author{R. Osborn}
\affiliation{Argonne National Laboratory, Argonne, IL 60439, USA}
\author{C. H. Wang}
\author{M. D. Lumsden}
\affiliation{Oak Ridge National Laboratory, Oak Ridge, TN 37831, USA}
\author{M. A. McGuire}
\author{A. S. Sefat}
\author{B. C. Sales}
\author{D. Mandrus}
\affiliation{Oak Ridge National Laboratory, Oak Ridge, TN 37831, USA}
\author{H. M. R\o nnow}
\affiliation{Laboratory for Quantum Magnetism, Ecole Polytechnique F\'ed\'erale de Lausanne, CH-1015, Switzerland}
\author{Y. Su}
\affiliation{J\"ulich Centre for Neutron Science, FZ J\"ulich, Outstation at FRM II, D-85747 Garching, Germany}
\author{A. D. Christianson}
\affiliation{Oak Ridge National Laboratory, Oak Ridge, TN 37831, USA}
\email{ROsborn@anl.gov}

\begin{abstract}
We report inelastic neutron scattering measurements of crystal field transitions in PrFeAsO, \PFAOF, and \NFAOF. Doping with fluorine produces additional crystal field excitations, providing evidence that there are two distinct charge environments around the rare earth ions, with probabilities that are consistent with a random distribution of dopants on the oxygen sites. The 4$f$ electrons of the Pr$^{3+}$ and Nd$^{3+}$ ions have non-magnetic and magnetic ground states, respectively, indicating that the enhancement of  \Tc\ compared to \LFAOFx\ is not due to rare earth magnetism.
\end{abstract}

\date{\today}

\maketitle

Tuning materials to enhance their properties is the driving force behind much of modern condensed matter physics.  Chemical doping is often the most practical means of accomplishing this goal but a thorough microscopic understanding of the effect of doping is a major challenge.  For example, the role of phase separation in the high-temperature superconducting cuprates\cite{Muller:1993p1} and colossal magnetoresistive manganites\cite{Moreo:1999p2841} is still being debated.  In the recently discovered iron-based superconductors, chemical doping is the primary method of inducing superconductivity, but there is conflicting evidence whether this is due to chemical pressure\cite{Takahashi:2008p7382}, the change in carrier concentration\cite{Wen:2008p7965}, disorder\cite{Wadati:2010p34986}, or a combination of all three.  Moreover, the question of phase separation in the superconducting phase itself has not been conclusively answered and may be material-dependent, with evidence of both phase separation into antiferromagnetic and superconducting regions\cite{Park:2009p20194} and phase coexistence\cite{Fernandes:2010p33898}. 

The \RFAOFx\ ($R=$La, Ce, Pr, Nd, Sm, Gd) series\cite{Kamihara:2008p7994,Ren:2008p7759,Ren:2008p35719,Ren:2008p35740,Chen:2008p8161,ChengPeng:2008p35965,Kadowaki:2009p36063,Khlybov:2009p36031} was the first family of iron-based superconductors to be discovered. Replacement of lanthanum with other rare earths increases \Tc\ up to $\sim 55$\,K in the optimally doped regime close to $x=0.15$.  Doping with fluorine adds electrons into iron $d$-bands but the substitution of trivalent rare earths for lanthanum does not change the carrier concentration, so any changes are either due to the influence of local 4$f$ magnetism or the effect of chemical pressure from the well-known lanthanide contraction.  Understanding which of these are responsible for the enhancement of \Tc\ has not been resolved.  Recent neutron diffraction, muon spin relaxation ($\mu$SR), and M\"ossbauer studies of the $R$FeAsO parent compounds have shown evidence for strong coupling between rare earth and iron magnetism\cite{Kimber:2008p35766,Zhao:2008p13387,Maeter:2009p30898,McGuire:2009p33606}. Assuming that this coupling persists in fluorine-doped systems, we might expect the 4$f$ moments to influence the superconducting properties as well. Indeed, a recent $^{19}$F NMR study of \SFAOFx has inferred a non-negligible coupling between the 4$f$ and conduction electrons\cite{Prando:2010p35776}.

Measurements of rare earth crystal field (CF) excitations using inelastic neutron scattering provide unique insight into the effects both of substituting magnetic rare earth ions for lanthanum and of doping fluorine onto the oxygen sublattice. This is because it is both a local probe, since the CF potential acting on the 4$f$ electrons is determined by the electrostatic environment produced predominantly by the nearest-neighbor oxygen/fluorine ions, and a bulk probe, since the CF transition intensities are a true thermodynamic average of the whole sample. Relative changes in the CF peak intensities can be directly related to the volume fraction of rare earth ions affected by a particular configuration of neighboring ligands.

In this article, we present a comparison of CF excitations measured in \PFAOFx with $x = 0.0$ and 0.13, which shows that there are two distinct charge environments in the superconducting compound, similar to the conclusions of a recent $^{75}$As NQR study\cite{Lang:2010p32185}. However, the measured reduction in CF intensity with $x$ is consistent with a random distribution of fluorine ions, showing that the charge environments are produced by the dopant ions and not electronic phase separation in the iron layers, as proposed in the earlier work.  Similarly, our measurements on \NFAOF\ reveal more crystal field excitations than would be allowed due to either tetragonal or orthorhombic symmetry, also indicating the presence of inequivalent rare earth sites. Moreover, we are able to eliminate or severely constrain the relevance of rare earth magnetism to the pairing mechanism since superconducting \PFAOF\ has a singlet ground state while superconducting \NFAOF\ has a magnetic Kramer's doublet ground state, even though both the superconducting transition temperatures are similar\cite{Ren:2008p7759,Ren:2008p35719}.  
   
Powder samples of \RFAOFx were synthesized following the method described in Ref. \cite{McGuire:2009p33606}. Superconducting transition temperatures determined by the onset of diamagnetism in an applied field of 20\,Oe are 41\,K and 49\,K for \PFAOFx\ ($x=0.13\pm0.01$ determined from the phase diagram of Ref. \cite{Rotundu:2009p36139}), and NdFeAsOF$_{0.15}$ (nominal composition), respectively. Structural characterization by neutron diffraction was performed on GEM at the ISIS Facility and HB2A at the High Flux Isotope Reactor. Inelastic neutron scattering (INS) studies were conducted on time-of-flight spectrometers Merlin at ISIS and IN4 and IN6 at the Institut Laue Langevin. The INS data have been placed on an absolute scale by normalization to a vanadium standard.

\begin{figure}[!b]
\centering
\vspace{-0.2in}
\includegraphics[width=\columnwidth]{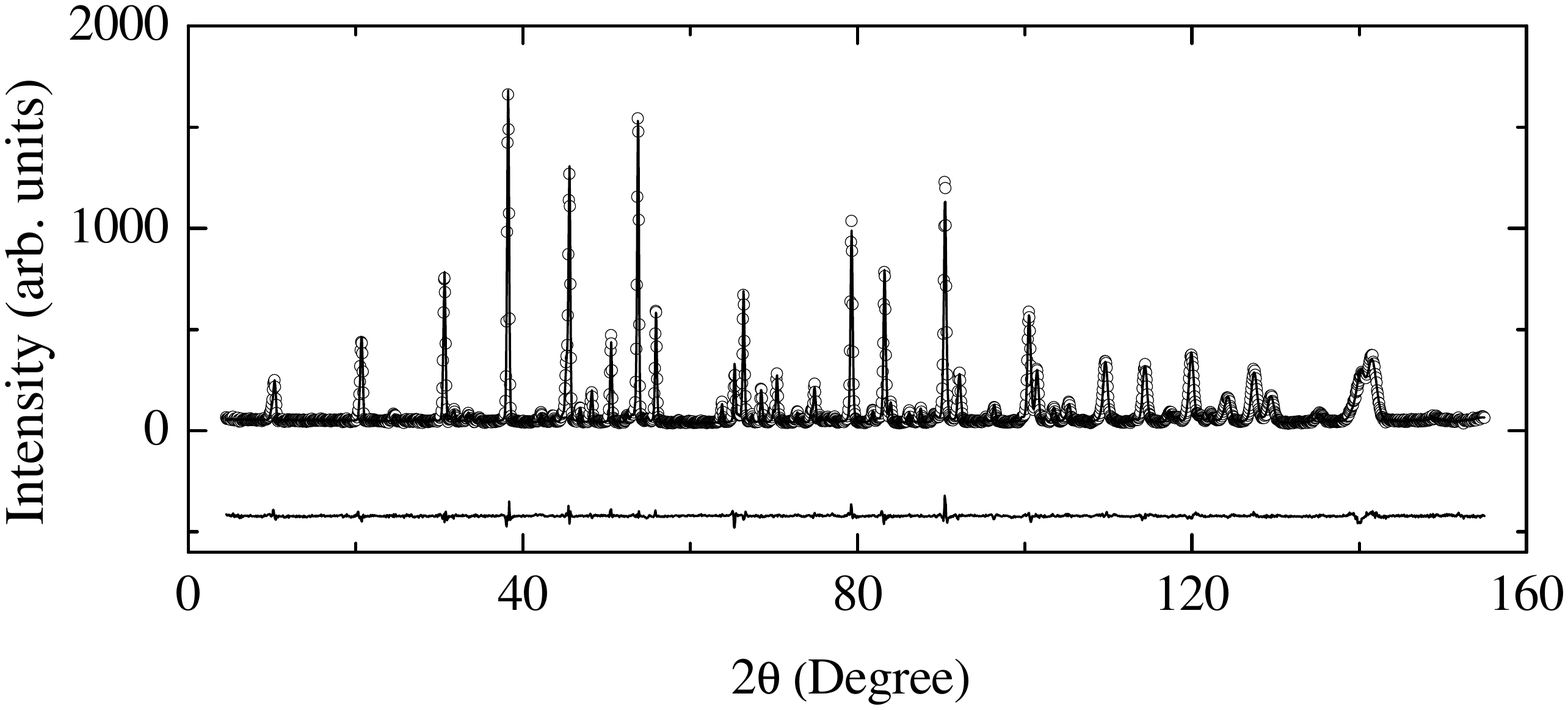}
\caption{Neutron powder diffraction in \PFAOF\ measured on HB2A at 4\,K (circles) compared to the Rietveld refinement (line). The line below the plot shows the difference between the data and the refinement. 
\label{Fig1} }
\vspace{-0.3in}
\end{figure}

Fig. 1 shows the neutron diffraction pattern for \PFAOF\  measured at 4\,K, compared to a Rietveld fit using FullProf\cite{RodriguezCarvajal:1993p35834}. Structural refinements confirmed the orthorhombic (\textit{Cmma}) structure in the parent compound and the tetragonal ($P4/nmm$) structure in the superconducting sample, with evidence for $4.3\pm0.2$\% of FeAs impurity in both the $x=0.0$ and 0.13 compounds.  A PrOF impurity phase ($\sim3.8$\%) is also detectable in \PFAOF, but is too small to produce the CF excitations discussed below. There is no evidence of structural inhomogeneity in the primary phases of PrFeAsO and \PFAOF. No impurity phase was observed in \NFAOF.

Fig. 2 shows the CF excitations in PrFeAsO and \PFAOF\ below 14\,meV. The crystal field excitations are evident as the peaks at nonzero energy transfers that are not present in LaFeAsO, which has no 4$f$ electrons,   There are additional CF excitations between 30 and 40\, meV, which are not shown.   In the parent compound, PrFeAsO, the CF excitations centered at $\sim3.5$\,meV are split above \TN(Pr)=12\,K. We assume that this is due to the internal molecular field created by the Fe sublattice below \TN(Fe) = 127\,K\cite{Zhao:2008p13387}, although we cannot confirm that they become degenerate above  \TN(Fe) because of thermal broadening.  We have insufficent information to solve the CF potential so we cannot construct a microscopic model to explain the collapse of the splitting when the rare earth sublattice magnetically orders as seen in the inset of Fig. 2(a). However, it may be due to a spin reorientation, similar to what has been observed in NdFeAsO\cite{Tian:2010p34291}, and is consistent with other evidence of an interplay between iron and rare earth magnetism in PrFeAsO, such as a reduction in the intensity of the iron magnetic Bragg peak\cite{Kimber:2008p35766} and a reduction in the $\mu$SR frequency\cite{Maeter:2009p30898} approaching  \TN(Pr).   

\begin{figure}[!b]
\centering
\vspace{-0.35in}
\includegraphics[width=0.7\columnwidth]{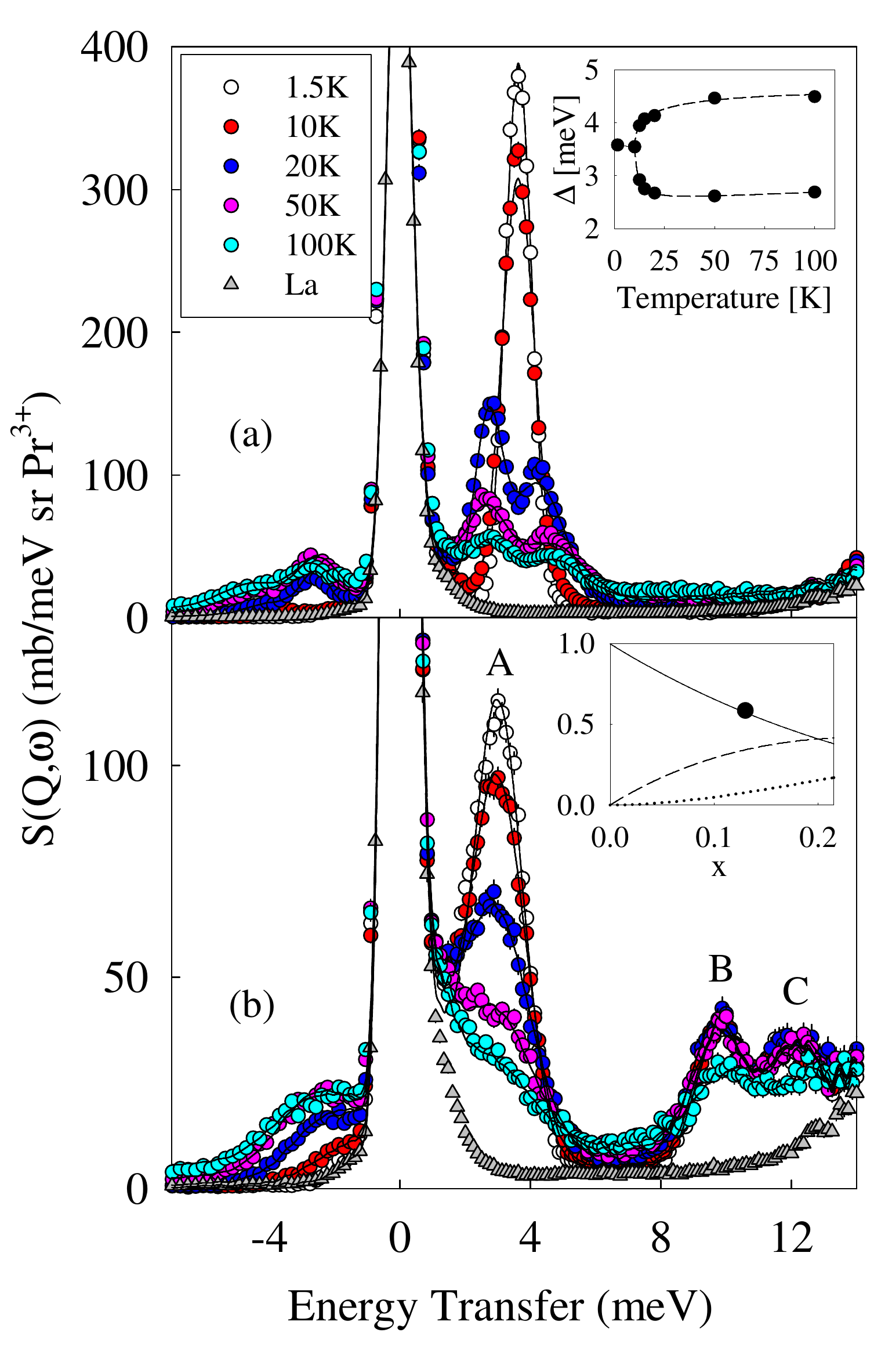}
\caption{(Color) Inelastic neutron scattering spectra of (a) PrFeAsO and (b) \PFAOF (circles) measured on IN4 with an incident energy of 17\,meV and an average elastic wavevector, \textit{Q}, of 0.85\,\AA$^{-1}$. The non-magnetic background is approximately given by the spectra of LaFeAsO (triangles). The solid lines in the panels (a) and (b) are fits to Gaussian lineshapes convolved with the instrumental resolution. (a) In PrFeAsO, the CF excitations from the ground state are centered at $3.58\pm0.01$\,meV at 1.5\,K. The inset shows the temperature evolution of the CF excitation energies. (b) In \PFAOF, A, B and C label CF excitations from the ground state, with energies of $2.78\pm0.01$\,meV, $9.72\pm0.05$\,meV, and $11.8\pm0.1$\,meV, respectively, measured at 1.5\,K. The inset shows the probabilities as a function of doping of the rare earth site having zero (solid line), one (dashed line), or two (dotted line) fluorine ions as nearest neighbors on the oxygen sublattice assuming a random distribution. The solid circle is the ratio of the intensities of the $\sim3$\,meV CF excitation in \PFAOF\ and PrFeAsO.
\label{Fig2} }
\vspace{-0.3in}
\end{figure}

In \PFAOF, there is also a CF peak at $\sim3$\,meV (labeled A) but there is no evidence of any splitting, which is consistent with the absence of long range magnetic order of the iron sublattice. However, the most striking observation is the appearance of two extra CF peaks at $9.72\pm0.05$\,meV and $11.8\pm0.1$ (labelled B and C, respectively)\,meV in the superconducting compound. There are no structural or magnetic phase transitions and the amount of PrOF impurity phase is far too small to explain them. The temperature dependence of their intensities confirm that all the transitions represent transitions from the ground state. However, the intensity of peak A decreases much faster than the intensities of B and C. As seen in Fig. 2(b), between 1.5\,K and 50\,K, the intensity of peak A falls by a factor of four whereas the intensities of peaks B and C remain almost the same. 

In order to quantify this observation, we have fit the measured data with a set of three Gaussian peaks and converted them to static susceptibilities in absolute units, using the Kramers-Kronig relations. The results of the temperature dependence of these fits are presented in Fig. 3. The susceptibilities of both the A and B transitions show typical Van Vleck behavior but their different temperature dependencies unambiguously indicate that these transitions belongs to two different rare-earth sites with different CF potential or charge environments.

\begin{figure}[!t]
\centering
\includegraphics[width=0.75\columnwidth]{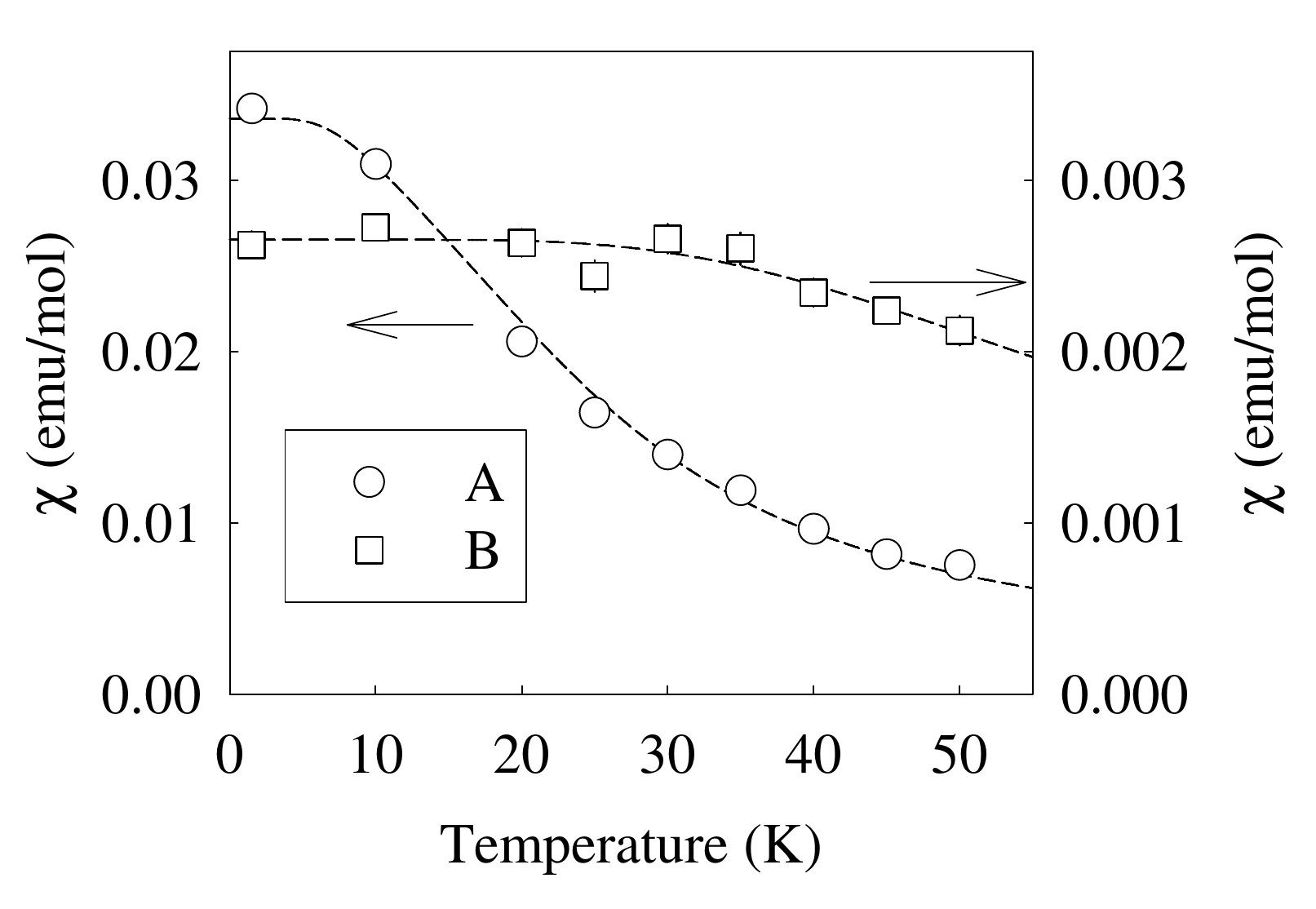}
\caption{Temperature dependence of the Van Vleck susceptibilities derived from the inelastic neutron scattering data for the crystal field transitions at $\sim 3.5$\,meV (A) and $\sim 9.7$\,meV (B). The temperature dependence of the intensity of peak C is similar to peak B.  
\label{Fig3} }
\vspace{-0.25in}
\end{figure}

The wide separation between the peaks at A and B show that fluorine doping strongly affects the local electrostatic potential felt by the rare earth ions.  Given that there are four oxygen nearest neighbors to the rare earth, up to five different crystal field spectra are possible due to configurations ranging from zero to four fluorine nearest neighbors.  Assuming a random distribution of fluorine atoms at this doping level would result in 57\% of the rare earth sites with no fluorine nearest neighbors.  Additionally 34\% of the rare earth sites would have one fluorine nearest neighbor.  The remaining 9\% of the rare earth sites would have two, three, or four fluorine nearest neighbors with decreasing probability.  Any tendency of the fluorine ions to cluster would alter these ratios. In particular, it would increase the fraction of rare earth sites with no fluorine ions and redistribute the remaining ratios. In fact, a comparison of the spectral weight of the peak at $\sim 3.5$\,meV between the superconducting compound and the parent compound yields a ratio of $58\pm4$\%, which is consistent with a random distribution fluorine dopants. Although it should be confirmed by measuring this ratio as a function of $x$, this value is sufficiently precise to be strong evidence against any substantial clustering of fluorine ions on the oxygen sublattice.

The widths of the CF transitions in both the parent and superconducting compounds are quite small and have a Gaussian line shape. The $\sim3$\,meV peak width (FWHM) at 1.5\,K is $0.61\pm0.02$\,meV in PrFeAsO and $1.54\pm0.03$\,meV in \PFAOF, after correction for the instrumental resolution. This indicates that there is some broadening from chemical disorder, probably produced by lattice strains due to longer-range fluctuations in the configuration of fluorine neighbors. The fact that we are seeing well defined, sharp CF transitions in the superconducting compound is evidence that the distribution of structural or electronic defects around the praseodymium sites is small and that we are dealing with two well-defined charge environments.     

\begin{figure}[!b]
\centering
\vspace{-0.3in}
\includegraphics[width=0.75\columnwidth]{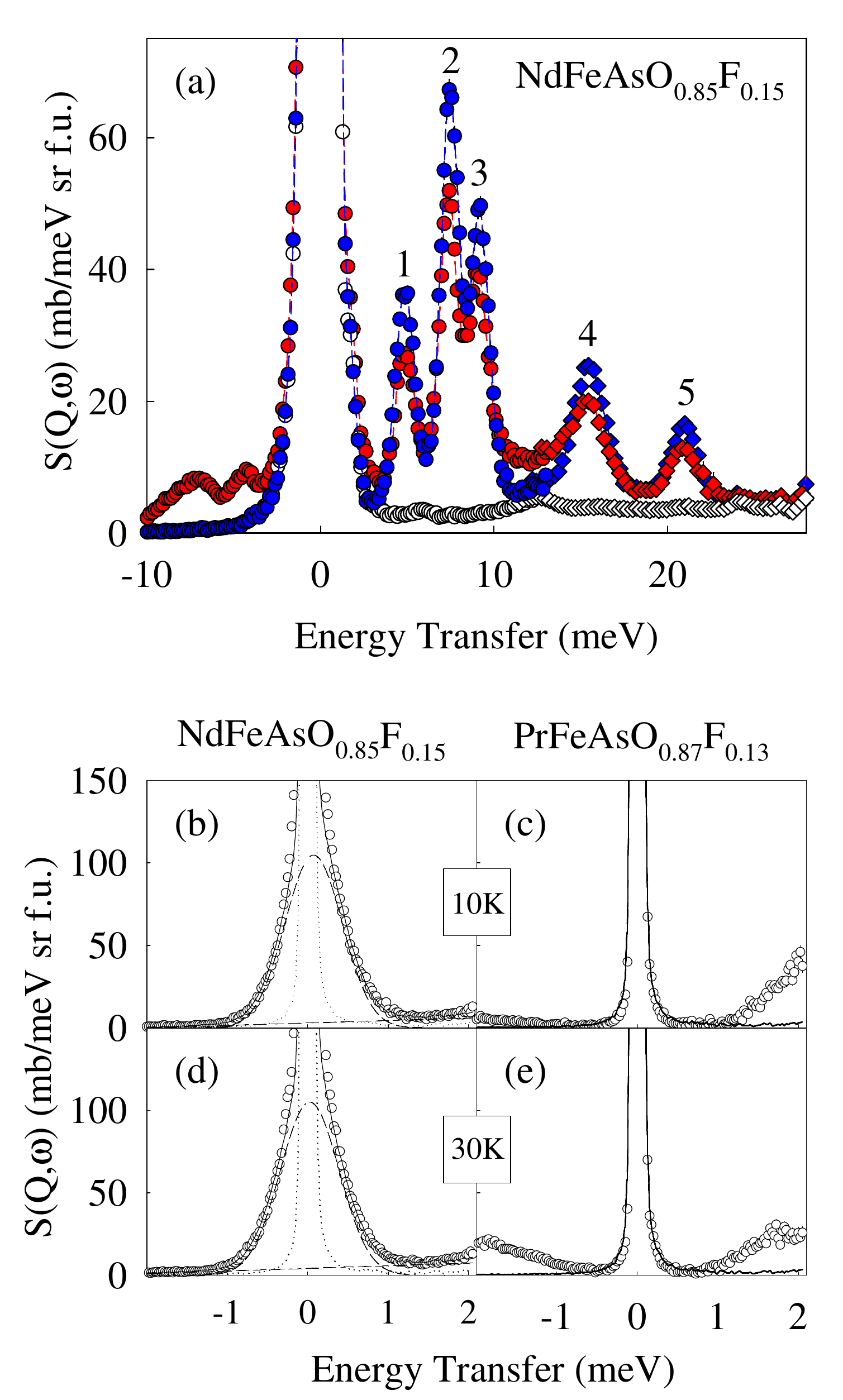}
\caption{(Color) (a) Inelastic neutron scattering spectra of \NFAOF measured on Merlin with incident energies of 15\,meV (circles) and 35\,meV (diamonds) at an average \textit{Q} of 1\,\AA$^{-1}$ and temperatures of 7\,K (blue symbols) and 50\,K (red symbols). Open symbols are non-magnetic scattering from \LFAOF. In the panels (b)-(e) the inelastic neutron scattering data measured on the IN6 with an incident energy of 3.1\,meV and an average elastic \textit{Q} of 0.65\,\AA$^{-1}$ for \NFAOF (b), (d) and \PFAOF (c), (e) at 10\,K (b), (c) and 30\,K (d), (e). The dotted line in (b), (d) and solid line in (c), (e) is the elastic nuclear scattering. In (b) and (d), the dashed line is the quasielastic Gaussian line shape fit and the solid line is the sum of elastic nuclear, magnetic and small linear background contributions. 
\label{Fig4} }
\end{figure}

Another example where fluorine doping produces multiple rare earth environments in a \RFAOF material is for $R=$Nd. The INS data for \NFAOF measured on the Merlin spectrometer at 7\,K and 50\,K and at initial energies 15\,meV and 30\,meV are shown in Fig. 4(a).  For Nd$^{3+}$ ions in a point group symmetry lower than cubic, \textit{i.e.} tetragonal, orthorhombic, or monoclinic, the $J=9/2$ ground state multiplet breaks up into five Kramers doublets, so there can be at most four CF transitions from the ground state. There are clearly five peaks due to ground state CF transitions and thus, while it is difficult to assign each peak to a particular rare earth site, the data are consistent with the picture presented above, in which the local electrostatic potential of the rare earth ion is modified by fluorine doping. 

These observations bear some similarity to recent NQR measurements showing the existence of two charge environments in underdoped \RFAOF for $R=$La and Sm\cite{Lang:2010p32185}. $^{75}$As NQR is sensitive to the the electric field gradient, \textit{i.e.}, the second-degree CF potential, acting on the arsenic sites, which is also affected by changes in the local fluorine ion distribution. However, Lang \textit{et al} argue that the two spectral components they observe in the superconducting phase is due to local electronic order on the iron layer, because of the $x$-dependence of the spectral weights. It is more difficult to model the effect of different fluorine ion configurations because of the greater distance of the arsenic ions from the oxygen/fluorine sublattice and the longer-range of the second-degree CF potential, at least in ionic environments. It is clear that their data is more affected by inhomogeneous broadening than the neutron data making the spectral weight ratios more difficult to determine, so we cannot rule out their interpretation, but it seems unlikely that two such similar probes should both find evidence of two charge environments with completely different origins.

The final issue we wish to comment on is the effect that rare earth substitution has on superconductivity. As stated earlier, the fact that the rare earth substituents are isovalent to lanthanum mean that the increase in \Tc\ is either due to chemical pressure from their smaller ionic size or due to the 4$f$ magnetic moments coupling to the iron magnetism and influencing the pairing interaction.  The high resolution neutron scattering data shown in Fig 4(b-e) show that the low energy magnetic fluctuations are very different in \PFAOF\ and \NFAOF.  In the case of neodymium, there is strong quasielastic scattering indicating a magnetic ground state as expected for a system of Kramer's doublets. On the other hand, the praseodymium sample exhibits no magnetic signal in this energy window, indicating unambiguously a nonmagnetic singlet ground state.  Therefore, even though the superconducting transition temperature is nearly the same for both materials at optimal doping, the rare earth ground states are very different in both materials. This makes it unlikely that the 4$f$ magnetic moment is involved in the superconductivity in these materials.  Thus, we conclude that it is the effect of chemical pressure due to the lanthanide contraction that is responsible for enhancing \Tc\ when lanthanum is replaced by another rare earth element.

In conclusion, we have measured the crystal field excitations in \RFAOF\ using inelastic neutron scattering and established the existence of two charge environments for rare earth sites in nearly optimally-doped \PFAOFx\  and \NFAOFx compounds that are due to a random distribution of fluorine ions, rather than electronic phase separation on the iron layers as proposed in an earlier NQR study\cite{Lang:2010p32185}. Measurements of low-energy magnetic fluctuations reveal that Pr$^{3+}$ and Nd$^{3+}$ 4$f$ elecrons have nonmagnetic and magnetic CF ground states, respectively, from which we infer that the 4$f$ magnetic moments are not responsible for the nearly identical enhancement of superconductivity in these compounds compared to \LFAOFx\ and conclude that a more likely candidate is chemical pressure produced by the lanthanide contraction.   

We acknowledge useful discussions with E. Dagotto and assistance in the neutron scattering experiments from O. Garlea (ORNL), T. Guidi (ISIS), A. Orecchini (ILL) and M. Koza (ILL). Research at Argonne and Oak Ridge is supported by the U.S. Department of Energy, Office of Science, Office of Basic Energy Sciences, Materials Sciences and Engineering Division and Scientific User Facilities Division.


\begin{thebibliography}{10}

\bibitem{Muller:1993p1}
 {\em Phase Separation in Cuprate Superconductors}, edited by K. M\"uller
  and G. Benedek (World Scientific Pub Co Inc., Italy, 1993).

\bibitem{Moreo:1999p2841}
A. Moreo, S. Yunoki, and E. Dagotto, Science {\bf 283},  2034  (1999).

\bibitem{Takahashi:2008p7382}
H. Takahashi {\it et~al.}, Nature {\bf 453},  376  (2008).

\bibitem{Wen:2008p7965}
H.-H. Wen {\it et~al.}, EPL {\bf 82},  17009  (2008), rB820311.

\bibitem{Wadati:2010p34986}
H. Wadati, I. Elfimov, and G.~A. Sawatzky, Phys Rev Lett {\bf 105},  157004
  (2010).

\bibitem{Park:2009p20194}
J.~T. Park {\it et~al.}, Phys Rev Lett {\bf 102},  117006  (2009).

\bibitem{Fernandes:2010p33898}
R.~M. Fernandes and J. Schmalian, Phys Rev B {\bf 82},  014521  (2010).

\bibitem{Kamihara:2008p7994}
Y. Kamihara, T. Watanabe, M. Hirano, and H. Hosono, J Am Chem Soc {\bf 130},
  3296  (2008).

\bibitem{Ren:2008p7759}
Z.-A. Ren {\it et~al.}, EPL {\bf 82},  57002  (2008).

\bibitem{Ren:2008p35719}
Z.-A. Ren {\it et~al.}, Mater Res Innov {\bf 12},  105  (2008).

\bibitem{Ren:2008p35740}
Z.-A. Ren {\it et~al.}, Chinese Phys Lett {\bf 25},  2215  (2008).

\bibitem{Chen:2008p8161}
G.~F. Chen {\it et~al.}, Phys Rev Lett {\bf 100},  247002  (2008).

\bibitem{ChengPeng:2008p35965}
C. Peng {\it et~al.}, Sci China Ser G {\bf 51},  719  (2008).

\bibitem{Kadowaki:2009p36063}
K. Kadowaki, A. Goya, T. Mochiji, and S. Chong, Journal of Physics: Conference
  Series {\bf 150},  052088  (2009).

\bibitem{Khlybov:2009p36031}
E.~P. Khlybov {\it et~al.}, JETP Lett {\bf 90},  387  (2009).

\bibitem{Kimber:2008p35766}
S.~A.~J. Kimber {\it et~al.}, Phys. Rev. B {\bf 78},  140503  (2008).

\bibitem{Zhao:2008p13387}
J. Zhao {\it et~al.}, Phys Rev B {\bf 78},  132504  (2008).

\bibitem{Maeter:2009p30898}
H. Maeter {\it et~al.}, Phys Rev B {\bf 80},  094524  (2009).

\bibitem{McGuire:2009p33606}
M.~A. McGuire {\it et~al.}, New J Phys {\bf 11},  025011  (2009).

\bibitem{Prando:2010p35776}
G. Prando {\it et~al.}, Phys. Rev. B {\bf 81},  100508  (2010).

\bibitem{Lang:2010p32185}
G. Lang {\it et~al.}, Phys Rev Lett {\bf 104},  097001  (2010).

\bibitem{Rotundu:2009p36139}
C. Rotundu {\it et~al.}, Phys. Rev. B {\bf 80},  144517  (2009).

\bibitem{RodriguezCarvajal:1993p35834}
J. Rodr{\'\i}guez-Carvajal, Physica B {\bf 192},  55  (1993).

\bibitem{Tian:2010p34291}
W. Tian {\it et~al.}, Phys Rev B {\bf 82},  060514  (2010).

\end{thebibliography}

\end{document}